\documentclass[11pt,a4paper]{article}
\usepackage[utf8]{inputenc}
\usepackage[margin=1in]{geometry}
\usepackage{amsmath}
\usepackage{amssymb}
\usepackage{graphicx}
\usepackage{hyperref}
\usepackage{enumitem}
\usepackage{authblk}
\usepackage{natbib}
\usepackage{url}
\usepackage{xcolor}
\usepackage{listings}

\definecolor{codegreen}{rgb}{0,0.6,0}
\definecolor{codegray}{rgb}{0.5,0.5,0.5}
\definecolor{codepurple}{rgb}{0.58,0,0.82}
\definecolor{backcolour}{rgb}{0.95,0.95,0.92}

\lstdefinestyle{mystyle}{
    backgroundcolor=\color{backcolour},
    commentstyle=\color{codegreen},
    keywordstyle=\color{magenta},
    numberstyle=\tiny\color{codegray},
    stringstyle=\color{codepurple},
    basicstyle=\ttfamily\footnotesize,
    breakatwhitespace=false,
    breaklines=true,
    captionpos=b,
    keepspaces=true,
    numbers=left,
    numbersep=5pt,
    showspaces=false,
    showstringspaces=false,
    showtabs=false,
    tabsize=2
}

\title{Verifiable Off-Chain Governance}

\author[1]{Jake Hartnell}
\author[2]{Eugenio (Neno) Battaglia}

\affil[1]{wavs.xyz}
\affil[2]{meaning.systems}

\date{December 2024}

\begin{document}

\maketitle

\begin{abstract}
Current DAO governance praxis limits organizational expressivity and reduces complex organizational decisions to token-weighted voting due to on-chain computational limits. This paper proposes verifiable off-chain computation (leveraging Verifiable Services, TEEs, and ZK proofs) as a framework to transcend these constraints while maintaining cryptoeconomic security. This paper explores three novel governance mechanisms: (1) attestation-based systems that compute multi-dimensional stakeholder legitimacy, (2) collective intelligence through verifiable preference processing, and (3) autonomous policy execution via Policy-as-Code. The framework provides architectural specifications, security models, and implementation considerations for DAOs seeking higher-resolution expressivity and increased operational efficiency, with validation from pioneering implementations demonstrating practical viability.
\end{abstract}

\section{Introduction}

Decentralized Autonomous Organizations promised to rival traditional organizational dynamics through game-theoretic coordination mechanisms that align incentives at scale. These mechanisms leverage the cryptoeconomic and governance primitives enabled by blockchain's inherent transparency and trustless execution. However, after a decade of experimentation, DAO governance remains primitive, often reducing complex organizational decisions to unidimensional token-weighted voting on predetermined proposals \cite{buterin2021moving}.

Smart contract execution environments constrain DAOs to their current primitive state. They cannot translate semantically rich off-chain contributions into on-chain governance decisions, nor implement liquid democracy with expertise-based delegation, nor execute adaptive policies that respond to changing conditions, nor can they coordinate effectively across blockchains. The same constraints that granted blockchain's security and verifiability now bottleneck DAOs from developing the multi-dimensional governance mechanisms complex coordination requires.

Verifiable Off-chain Services (hereafter ``Verifiable Services'') and Zero Knowledge proofs enable a technical breakthrough that unlocks decentralized governance's full speed and resolving power through cryptoeconomically secured off-chain computation. This paper presents a framework enabling three fundamental categories of governance innovation:

\begin{enumerate}
\item \textbf{A formal model for attestation-based governance} that captures multi-dimensional stakeholder legitimacy computing voting weights from verifiable social graphs. This enables sophisticated delegation, value-based incentives, and cross-DAO coordination.

\item \textbf{Collective intelligence mechanisms} that process semantically rich community preferences through fixed-stack, auditable DAO to Language Model pipelines embedded in the governance engine. These pipelines increase the throughput and resolution of governance by accelerating the pace and granularity of collective decision-making.

\item \textbf{A Policy-as-Code execution framework} that enables continuous organizational operations while maintaining democratic oversight.
\end{enumerate}

For instance, attestation graphs counterintuitively bolster Sybil resistance by converting ``social noise'' into verifiable legitimacy signals, while Policy-as-Code amplifies human oversight by making automated decisions auditable and contestable, transforming plutocratic weaknesses into meritocratic strengths.

Unlike oracles that merely import external data, this model synthesizes attestations and preferences into verifiable policies without trusted intermediaries, addressing pain points like cross-chain fragmentation through seamless federation. By processing rich, verifiable inputs at scale, it mitigates risks such as attestation privacy breaches and new capture forms via clear trust boundaries, with off-chain computations verified economically or cryptographically on-chain.

Ultimately, this framework equips DAOs to match or exceed legacy organizations' efficiency in throughput and reliability, while preserving transparency, inclusivity, and capture resistance. Rather than constraining institutions to smart contract limits, DAOs can build governance systems tailored to 21st-century coordination demands.

\section{The Governance Bottleneck: Smart Contracts}

Smart contracts deliver public verifiability and credible execution \cite{szabo1997formalizing}. They also impose hard limits that shape what governance can do. For instance, a blockchain's gas limit restricts the complexity of operations that can be performed in a single transaction. While these limitations are essential for blockchain security, they create significant barriers to implementing sophisticated governance mechanisms. Three constraints dominate in practice: computation is costly, data is siloed, and execution is synchronous. This section details each constraint and ties it to real governance needs.

\subsection{Computational Constraints and Their Governance Implications}

On-chain computation is bounded by per-transaction gas budgets and conservative execution rules. Complex algorithms become impractical once a community grows.

Liquid democracy shows the problem clearly. Effective voting power requires resolving multi-hop delegations, detecting cycles, and aggregating weights across the reachable subgraph for each voter or proposal. The work scales with the number of delegations and the depth of chains. At DAO scale, this traversal does not fit within a safe transaction. Splitting the computation across many transactions introduces race conditions.

In Compound, more than 165,000 token holders and deep delegation chains make exact on-chain resolution unrealistic. The implementation falls back to simplified models that fit the budget. The outcomes match the constraint. Participation by delegated COMP hovered near 20 percent in 2024 and a small set of about 13 delegates could steer a majority. The mechanism did not begin that way. It was pushed there by cost.

The same ceiling appears in other governance computations. Rich preference aggregation such as Condorcet or STV requires repeated pairwise comparisons and elimination rounds. Quadratic voting requires per identity squaring with identity control. Eligibility checks across multiple token positions and historical contribution windows require repeated cross-references. Any of these can be done for a few voters. They do not scale on chain without degradation.

Storage and iteration costs matter too. Counting over large token holder sets forces repeated reads of account balances or snapshot structures. Even when data exists on chain, scanning it during a vote inflates cost and exposes proposals to gas price volatility. Designers respond by narrowing the computation or by pushing work off chain without guarantees. Both routes reduce expressivity.

The implication is direct. Where the desired mechanism requires graph traversal, repeated comparison, or per voter personalization, gas replaces design as the binding constraint. DAOs then collapse toward unidimensional token voting and shallow quorum checks. Nuance is priced out.

\subsection{Data Isolation and Cross-Chain Fragmentation}

Smart contracts execute in closed environments. They do not read from external systems or other chains without extra machinery. Governance quality suffers when relevant signals live elsewhere.

Off-chain blind spots are common. Communities discuss and deliberate in forums, on GitHub, at events, and in working groups. Contracts cannot see that activity. MakerDAO is the clearest example. The forum shapes proposals through open discussion. Voting power is determined solely by MKR locked in the governance contract. A thoughtful contributor can influence opinions yet holds no formal weight unless they also hold tokens. This is a design choice driven by isolation. The contract cannot adjudicate forum reputation or review quality without a verifiable feed.

Cross-chain fragmentation compounds the problem. Token holders and activity are spread across Ethereum, Polygon, Arbitrum, Optimism, and other networks. Unifying governance today often means asking voters to bridge assets, wait through finality windows, and accept extra risk. Many abstain. Even simple tasks like computing a voter's aggregate stake across chains become awkward. Without a reliable way to view and combine positions, organizations either centralize voting on a single chain or accept fractured participation.

The result is governance that rewards assets over participation, lacks nuance in source legitimacy, and forces communities into uncomfortable centralization. Technical limits, not organizational preference, dictate the shape of coordination.

\subsection{Synchronous Execution and Timing Constraints}

Blockchain execution is strictly synchronous. Each state change follows the previous in a deterministic sequence. For payments and token transfers, this model provides essential ordering and safety. For governance workflows, it creates brittleness and inefficiency.

Operational governance needs asynchronous processing. A DAO might evaluate grant applications in parallel, screen proposals for quality before voting, or update treasury allocations based on continuous monitoring. These workflows involve multiple agents working concurrently with different timing requirements. On chain, they must be linearized into rigid sequences.

The cost in time and capital is severe. A grant review pipeline that could complete in days when parallelized gets stretched across weeks of sequential votes. Emergency responses wait behind routine operations. The DAO becomes unresponsive to time-sensitive opportunities.

Real governance also requires conditional logic that adapts to changing context. If market conditions shift, treasury allocations should adjust. If participation drops, quorum rules might flex. If a dependency fails, the workflow should handle the exception gracefully. Smart contracts struggle with this adaptability. Complex conditional paths multiply gas costs and introduce state explosion risks. Most implementations simplify to static rules that cannot respond to context.

\subsection{Requirements for Breakthrough}

These constraints suggest the requirements for a governance breakthrough:

\begin{itemize}
\item \textbf{Flexible computation} that can run complex algorithms (graph analysis, ML inference, optimization) without gas limits
\item \textbf{Rich data integration} that verifiably incorporates off-chain activity and cross-chain state
\item \textbf{Asynchronous workflows} with parallel processing, conditional logic, and adaptive timing
\item \textbf{Verification without re-execution} so contracts can trust results without repeating expensive work
\item \textbf{Graceful degradation} that maintains security even when components fail or actors misbehave
\end{itemize}

These requirements set the stage for the next sections. Verifiable Services and TEEs address flexible computation. Zero-knowledge proofs address standardized private or expensive checks. A hybrid architecture ties them to minimal on-chain interfaces that enforce verification, timelocks, and recourse.

\section{Verifiable Services: Enabling New Governance Designs}

This section describes three primitives that lift the limits set in Section 2 while keeping verifiability. Trusted Execution Environments provide hardware-based guarantees of correct execution. Verifiable Services compose multiple operators—potentially running TEEs—into economically secured networks. Zero-knowledge proofs handle fixed computations that benefit from strong privacy or strict determinism. Hybrid architectures may leverage the strengths of each approach, promoting maximum verifiability, autonomy, and resilience.

\subsection{Trusted Execution Environments}

A Trusted Execution Environment (TEE) is a hardware-isolated region of a processor that executes code in a protected enclave. The hardware ensures that even the machine's owner cannot observe or tamper with computation inside the enclave. Remote parties can verify what code is running through cryptographic attestation.

\subsubsection{Core Properties}

TEEs provide three guarantees relevant to governance:

\begin{enumerate}
\item \textbf{Integrity:} Code running inside the enclave cannot be modified. The hardware prevents external interference, and attestation proves the exact binary being executed.
\item \textbf{Confidentiality:} Data inside the enclave remains encrypted in memory. Neither the operating system nor other processes can read enclave contents.
\item \textbf{Attestation:} The TEE generates cryptographic proofs that specific code is running on genuine hardware. Remote verifiers can confirm execution environment integrity before trusting outputs.
\end{enumerate}

For governance, these properties enable a distinct trust model. Rather than trusting operators to behave honestly (economic security) or verifying computation through proofs (cryptographic security), participants trust that hardware correctly executes attested software.

\subsubsection{Trust Model}

TEE security rests on different foundations than economic or cryptographic approaches:

\begin{itemize}
\item \textbf{Hardware root of trust:} Security derives from processor design and manufacturing. The hardware enforces isolation regardless of software-level attacks.
\item \textbf{Attestation verification:} Before accepting results, verifiers check that outputs came from genuine hardware running expected code. Attestation chains link outputs to specific software versions.
\item \textbf{Manufacturer trust:} The security model assumes hardware manufacturers have not introduced backdoors and that attestation keys remain uncompromised.
\item \textbf{Side-channel awareness:} While enclaves protect against direct observation, sophisticated attacks can infer information through timing, power consumption, or memory access patterns. Implementations must account for these vectors.
\end{itemize}

A single TEE provides strong guarantees that software executed correctly, but concentrates trust in one hardware instance and its manufacturer. This creates a spectrum of deployment options with different trust assumptions.

\subsubsection{Governance Applications}

TEEs enable governance patterns that benefit from execution integrity:

\begin{itemize}
\item \textbf{Confidential preference aggregation:} Collect votes inside an enclave that tallies results without exposing individual choices. Attestation proves the aggregation algorithm ran correctly.
\item \textbf{Sealed computation:} Execute sensitive logic—compensation calculations, eligibility checks, scoring algorithms—where even operators should not see intermediate values.
\item \textbf{Key management:} Governance keys can live inside enclaves, signing transactions only when attested code approves them. This enables automated execution without exposing keys to operators.
\item \textbf{Deterministic randomness:} Generate verifiable random values inside enclaves for tie-breaking, selection, or sampling. The hardware prevents manipulation while attestation proves the generation method.
\end{itemize}

\subsection{Verifiable Services}

A Verifiable Service (Sometimes called ``Autonomous Verifiable Service'' or ``Network Extension'') is a set of bonded operators that run agreed software off-chain and return results that on-chain contracts can check and accept \cite{eigenlayer2023}. This approach builds on existing oracle designs while adding verifiable computation guarantees.

\subsubsection{Core Components}

The system has four parts:

\begin{enumerate}
\item \textbf{Operators:} The entities running the software, each bonded with stake that can be slashed for misbehavior
\item \textbf{Task specification:} The code to run, input requirements, and acceptance criteria encoded in a registry contract
\item \textbf{Execution and aggregation:} Operators fetch inputs, run tasks, and submit signed outputs
\item \textbf{Verification and settlement:} Contracts check outputs meet acceptance criteria, aggregate results, slash for failures
\end{enumerate}

For governance, operators might be established node runners, professional validators, or community-elected technical teams. Tasks could include preference aggregation, proposal quality scoring, or treasury rebalancing computations. The verification focuses on output consistency and timeliness rather than re-executing the computation.

\subsubsection{Trust Model}

Verifiable Services provide economic rather than cryptographic guarantees. Security comes from:

\begin{itemize}
\item \textbf{Multi-operator redundancy:} Multiple operators run the same task and must achieve threshold agreement (e.g., 3-of-5)
\item \textbf{Slashing for misbehavior:} Operators stake bonds that get slashed for submitting incorrect results or missing deadlines
\item \textbf{Deterministic software:} Tasks use fixed code versions, pinned dependencies, and deterministic execution to ensure reproducibility
\item \textbf{Auditability:} All inputs, outputs, and code are public, allowing anyone to verify correctness and dispute if needed
\end{itemize}

The economic security scales with the value at stake. For high-value governance decisions, larger bonds and more operators can be required. For routine operations, lighter requirements reduce cost.

\subsubsection{Example: Delegation Graph Resolution}

Consider liquid democracy with transitive delegation:

\begin{enumerate}
\item \textbf{Registry stores task:} ``Resolve delegation graph for proposal X using snapshot at block B''
\item \textbf{Operators fetch data:} Download delegation records and token balances at specified block
\item \textbf{Compute voting power:} Run graph traversal to resolve chains, detect cycles, aggregate weights
\item \textbf{Submit results:} Each operator signs and submits merkle root of voter weights
\item \textbf{Contract verifies:} Check 3-of-5 operators submitted matching roots within deadline
\item \textbf{Settlement:} Accept root, pay operators, slash any who submitted different results
\end{enumerate}

The computation that would cost millions in gas runs efficiently off-chain. The contract only verifies signatures and compares 32-byte roots. Voters can generate merkle proofs to claim their resolved voting power.

\subsection{Zero-Knowledge Proofs}

ZK proofs provide cryptographic guarantees that computation was performed correctly without revealing inputs or requiring re-execution. They are ideal for privacy-sensitive or standardized operations where the proof system's constraints are acceptable.

\subsubsection{Governance Applications}

Several governance patterns benefit from ZK:

\begin{itemize}
\item \textbf{Private voting:} Prove vote validity without revealing choice
\item \textbf{Eligibility checks:} Prove qualification (e.g., holding tokens, reputation score) without exposing exact amounts
\item \textbf{Aggregation:} Prove correct tallying of private inputs
\item \textbf{Compliance:} Prove proposals meet policy requirements without revealing sensitive details
\end{itemize}

For standardized computations with stable requirements, ZK provides stronger guarantees than economic security. The tradeoff is less flexibility—changing the circuit requires new trusted setup and deployment.

\subsubsection{Integration Points}

ZK proofs complement Verifiable Services:

\begin{itemize}
\item Services handle flexible, evolving logic
\item ZK handles fixed, privacy-sensitive operations
\item Services can generate and verify ZK proofs as part of larger workflows
\item Both feed minimal on-chain contracts that enforce governance decisions
\end{itemize}

\subsection{Hybrid Architecture}

The trust models—hardware, economic, and cryptographic—compose naturally, each addressing different failure modes. Production systems may combin multiple primitives.

The on-chain layer remains minimal. Contracts hold tokens, maintain registries of tasks and bonds, and verify outputs. Critical operations—asset transfers, permission changes—never leave smart contract control, protected by timelocks and dispute windows.

Off-chain, Verifiable Services handle complex computation: preference aggregation, proposal analysis, treasury rebalancing, cross-chain data collection. Operators can run inside TEEs for hardware integrity on top of economic bonds. TEEs also enable confidential computation—sensitive scoring algorithms, key custody, automated signing—where even operators cannot observe intermediate values. ZK proofs handle privacy-sensitive verification: private ballots, eligibility checks, compliance without disclosure.

The primitives reinforce each other. TEE-backed operators combine hardware and economic guarantees. Services may generate or verify ZK proofs as part of larger workflows. Economic bonds ensure liveness when cryptographic mechanisms ensure correctness.

If off-chain services fail, governance falls back to simple on-chain voting. The architecture degrades gracefully while preserving basic functionality.

The next sections show how verifiable offchain computation enables governance mechanisms previously impossible on-chain.

\section{Attestation-based Governance Systems}

Attestations are cryptographically signed statements that make verifiable claims about real-world attributes, relationships, or events \cite{w3c2022}. Unlike traditional credentials or certificates that rely on trusted authorities, attestations create a decentralized web of verifiable claims where anyone can attest to facts they have direct knowledge of. This includes confirming expertise in a domain, vouching for contributions to a project, or verifying participation in events.

While on-chain attestation protocols such as the Ethereum Attestation Service (EAS) \cite{ethereumattestationservice2023} have existed for a while, computation over attestation graphs in smart contracts was too expensive to leverage in DAO governance or incentive systems. Verifiable off-chain technologies change this by enabling verifiable computation over attestation graphs while keeping outputs checkable on chain (see Section 3). Networks like TrustGraph demonstrate this approach in production with 65+ experts participating in governance through attestation-based trust scores.

Attestation-based governance creates rich, verifiable social graphs that capture real-world attributes, relationships, and the nuanced reality of community contribution and expertise, rather than relying solely on token holdings or simple credential checks.

A smart contract can verify token ownership. It cannot assess whether someone has the expertise to evaluate a technical proposal, the relationships to broker a partnership, or the reputation to be trusted with treasury decisions. Attestations bridge this gap by enabling verifiable claims about these dimensions of stakeholder legitimacy.

\subsection{Architecture of Attestation-Based Governance}

Verifiable Services, a key off-chain computation technology, process attestation graphs through three functional layers:

\subsubsection{Attestation Infrastructure}

The data collection and verification layer is the primary infrastructure consisting of both on-chain attestation protocols and off-chain collection mechanisms:

\begin{itemize}
\item \textbf{On-chain attestation protocols:} Such as Ethereum Attestation Service (EAS), which provide persistent, cryptographically signed attestations that are publicly verifiable.
\item \textbf{Off-chain credential and peer-reviewed verification:} Systems that collect and verify credentials from external sources (academic credentials, GitHub contributions, professional certifications).
\item \textbf{Standardized claim schemas:} Defining what types of attestations exist and how they are structured for the specific governance context.
\item \textbf{Time-bound and revocable attestations:} Implementing expiry dates and revocation mechanisms to ensure attestations remain current and accurate.
\end{itemize}

\subsubsection{Computation Layer}

This is the key processing layer that performs expensive operations off-chain:

\begin{itemize}
\item \textbf{Graph traversal and analysis:} Computing influence scores, detecting communities, finding paths of trust.
\item \textbf{Multi-attribute scoring algorithms:} Weighing different types of attestations and contributions to generate composite legitimacy scores.
\item \textbf{PageRank-style trust propagation:} Calculating global trust scores based on the attestation graph structure.
\item \textbf{Social distance calculations:} Measuring degrees of separation for delegation limits or trust boundaries.
\item \textbf{Verifiable output generation:} Producing merkle roots, signatures, or ZK proofs that on-chain contracts can efficiently verify.
\end{itemize}

\subsubsection{Integration and Execution Layer}

This layer ties attestation insights to actual governance mechanisms:

\begin{itemize}
\item \textbf{Voting weight determination:} Using attestation scores to calculate voting power for proposals.
\item \textbf{Delegation routing:} Finding appropriate delegates based on expertise attestations.
\item \textbf{Access control:} Gating participation in specific decisions to attestation holders.
\item \textbf{Incentive distribution:} Allocating rewards based on verified contributions.
\item \textbf{Proposal routing:} Matching proposals to reviewers with relevant expertise attestations.
\end{itemize}

\subsubsection{Security Model}

The architecture implements multiple security layers:

\begin{itemize}
\item Attestations themselves are cryptographically signed and tamper-proof.
\item Verifiable Service operators stake bonds that can be slashed for incorrect computation.
\item Multiple operators must agree on computation results (e.g., 3-of-5 threshold).
\item On-chain contracts only accept results that meet verification criteria.
\item Time delays allow for disputes before high-stakes actions execute.
\end{itemize}

\subsubsection{Bootstrapping New Communities}

The cold-start problem is a critical challenge for attestation networks. New communities lack the dense web of attestations needed for meaningful trust scores. Implementations must design bootstrapping mechanisms that balance initial accessibility with long-term security.

One approach uses a three-layer bootstrapping strategy:

\textbf{Layer 1 - Genesis attestors:} A small set of widely trusted entities provide initial attestations. These could be established community members, respected organizations, or protocol founders. Their influence naturally decays over time through trust propagation algorithms.

\textbf{Layer 2 - Objective credentials:} External verifiable credentials provide initial legitimacy signals. This includes on-chain history (past participation in DAOs, protocol usage), off-chain credentials (GitHub contributions, professional certifications), or token holdings above minimum thresholds.

\textbf{Layer 3 - Progressive trust building:} New members earn attestations through small contributions initially, building reputation over time. Early participation is limited but grows with earned trust.

The bootstrapping configuration becomes a critical governance parameter. Too restrictive, and the network cannot grow. Too permissive, and Sybil attacks become feasible. Successful systems will likely require iteration and adjustment as they scale. Pioneering implementations like TrustGraph demonstrate viable solutions through trusted seeds and evidence-based onboarding.

\subsection{Use Cases}

\subsubsection{Liquid Democracy with Expertise-Weighted Delegation}

Token-based delegation systems enable basic vote delegation but cannot incorporate expertise or trust. A token holder can delegate to anyone, regardless of the delegate's knowledge or alignment with their interests. This creates risks of uninformed or misaligned decision-making at scale, where popular but unqualified delegates accumulate disproportionate power.

Attestation-based liquid democracy enables sophisticated delegation markets that consider multiple dimensions of delegate suitability:

\begin{itemize}
\item \textbf{Expertise matching:} Technical proposals route to delegates with verified technical expertise through attestations from peers who have worked with them
\item \textbf{Value alignment:} Delegators can restrict delegation to those who share attested values or have demonstrated commitment to specific causes
\item \textbf{Trust boundaries:} Delegation can be limited by social distance in the attestation graph, preventing delegation to unknown actors
\item \textbf{Context-specific delegation:} Different proposal types can route to different delegates based on relevant attestations
\item \textbf{Delegation markets:} Delegates can signal their expertise areas and capacity, while delegators can discover suitable representatives through attestation-based search
\end{itemize}

Verifiable Services compute these complex delegation paths by traversing the attestation graph, checking constraints, and resolving transitive delegations. The output is a merkle tree of resolved voting weights that the on-chain contract can verify without repeating the expensive computation. This enables DAOs to implement liquid democracy at scales previously impossible due to gas costs.

\subsubsection{Merit-Based Incentive Distribution}

Traditional DAOs distribute rewards based on simple metrics such as token holdings or proposal submission. This misses the vast majority of valuable contributions: thoughtful forum discussions, code reviews, community support, relationship building, and strategic guidance. These contributions are critical for DAO success but invisible to smart contracts. Emerging protocols like Hypercerts \cite{hypercerts2023} demonstrate growing interest in verifiable impact certification, though they focus on outcomes rather than the social graph of recognition.

Attestation-based merit systems recognize the full spectrum of contributions:

Contributors receive attestations from peers who directly observed their work. These might attest to code quality, helpful documentation, successful project management, or community building efforts. The attestations form a contribution graph showing who provided value and who recognized it.

Verifiable Services analyze this graph to compute contribution scores. The algorithm might weight recent contributions higher, recognize diverse contribution types, factor in the credibility of attestors, and detect and discount reciprocal attestation rings. The resulting scores determine reward distributions, creating incentives for the actual work that makes DAOs successful.

\subsubsection{Cross-DAO Reputation and Coordination}

As the DAO ecosystem matures, coordination between organizations becomes essential for shared challenges, common resources, and avoiding destructive competition. Traditional coordination relies on informal relationships or simple token-based voting across organizations. Attestation networks enable inter-organizational coordination that can scale while maintaining legitimacy.

Organizations attest to relationships with other DAOs, shared values, collaboration history, and trust levels. Individual members attest to participation in multiple organizations and to expertise in cross-organizational coordination. Verifiable services analyze these networks to identify natural coalition partners, detect conflicts of interest, and recommend coordination strategies aligned with each organization's stated goals and community preferences.

The system enables federated governance where related DAOs coordinate decision-making without surrendering autonomy. For ecosystem-wide challenges such as security standards, infrastructure funding, or regulatory responses, DAOs can participate in meta-governance where voting power reflects both organizational stake and attestation-based legitimacy. This avoids domination by large treasuries while reflecting genuine community preferences across organizations.

Cross-DAO attestation networks also enable reputation portability. Contributions to one organization can be recognized by others in the same ecosystem. This reduces friction for contributors who participate in multiple DAOs and creates incentives for collaboration.

\subsection{Theoretical Considerations for Implementation}

While attestation-based governance offers significant potential, implementations must address several theoretical challenges:

\textbf{Bootstrapping and Network Effects:} Cold-start presents a fundamental challenge—initial participants have no one to attest to them, while the value of attestations depends on network density. Implementations must design bootstrapping mechanisms that balance initial accessibility with long-term Sybil resistance. Pioneering implementations like TrustGraph demonstrate viable solutions through trusted seeds and evidence-based onboarding.

\textbf{Gaming and Coordination Risks:} Public attestation systems face potential manipulation through reciprocal attestation rings or social pressure dynamics. Decay mechanisms, negative attestations, and multi-dimensional scoring can mitigate but not eliminate these risks. The challenge lies in designing mechanisms robust to adversarial behavior while remaining accessible to legitimate participants.

\textbf{Privacy and Transparency Balance:} Attestations create permanent records that may reveal sensitive relationships or evaluations. While transparency enhances accountability, it may discourage honest negative feedback or expose professional networks. Future implementations might explore zero-knowledge attestations that prove properties without revealing specifics.

\textbf{Temporal Validity and Evolution:} Trust relationships change over time, requiring mechanisms for attestation expiry, updates, and revocation. Design choices around temporal validity significantly impact both system dynamics and computational requirements.

These considerations represent active areas of exploration rather than fundamental barriers. Each implementation will make different tradeoffs based on specific use cases and community values.

\subsection{Benefits of Attestation-based Governance Architecture}

Attestation-based governance architecture offers fundamental advantages over traditional DAO systems while enabling new organizational capabilities.

\begin{itemize}
\item \textbf{Multi-Dimensional Legitimacy:} Attestation networks recognize multiple legitimacy sources simultaneously such as domain expertise, contribution history, community trust, and context-specific knowledge.
\item \textbf{Dynamic Adaptability:} New contribution types can be recognized through new schemas without costly upgrades. Relative importance of factors can evolve based on community preferences.
\item \textbf{Scalable Quality Assurance:} Expertise recognition and peer validation improve as community size increases, combining crowd wisdom with expert influence.
\item \textbf{Sybil Resistance Through Social Structure:} Attestation requirements create natural barriers to fake accounts while allowing legitimate participants to build reputation over time.
\item \textbf{Contextual Flexibility:} Different decisions can weight attestations differently, enabling nuanced governance that adapts to decision type and importance.
\item \textbf{Permissionless Innovation:} Anyone can create new attestation types and scoring algorithms, enabling governance evolution without protocol changes.
\item \textbf{Interoperability:} Attestations can be recognized across different DAOs and chains, creating portable reputation and enabling ecosystem-wide coordination.
\end{itemize}

\section{Collective Intelligence through Verifiable Computation}

Traditional DAO voting compresses complex community preferences into binary yes/no decisions on pre-formed proposals. This compression discards valuable signal: the reasoning behind positions, conditions for support, suggested improvements, and expertise-weighted confidence. Communities actually possess rich collective intelligence. The limitation lies in aggregation mechanisms that can process only the simplest signals.

Verifiable Services enable high-bandwidth preference expression and sophisticated aggregation without sacrificing verifiability. This section explores how DAOs can tap their collective intelligence at unprecedented resolution.

\subsection{Semantically Rich Preference Expression}

Moving beyond binary voting requires accepting diverse input types while maintaining determinism. Citizens express preferences through multiple channels:

\textbf{Structured rubrics:} Multi-dimensional evaluation on defined criteria (feasibility, impact, alignment, risk) with weighted importance based on evaluator expertise.

\textbf{Preference rankings:} Ordering multiple options, expressing intensity of preference, and indicating acceptable alternatives for more nuanced collective decision-making.

\textbf{Conditional commitments:} Support contingent on specific modifications, resource availability, or concurrent decisions, enabling complex negotiation through governance.

\textbf{Resource allocation:} Distribution of budgets across proposals, projects, or teams based on perceived value, creating market-like signals within governance.

These inputs can be collected through standardized interfaces that ensure parseability while allowing expression flexibility. The key is defining clear schemas that balance expressiveness with computational tractability.

\subsection{Deterministic Processing Pipelines}

Verifiable Services process these rich inputs through deterministic pipelines. Every operation must be reproducible:

\textbf{1. Input validation and normalization:} Check signatures, verify eligibility, standardize formats, handle missing data consistently.

\textbf{2. Weight computation:} Calculate participant weights based on tokens, attestations, participation history, expertise in relevant domains.

\textbf{3. Aggregation logic:} Apply voting rules (quadratic, conviction, ranked choice), resolve conflicts and ties, compute confidence intervals.

\textbf{4. Output generation:} Produce canonical results, generate proof artifacts, create audit trails.

Critical: the pipeline must be deterministic. Given the same inputs, every operator must produce identical outputs. This requires:

\begin{itemize}
\item Fixed software versions with pinned dependencies
\item Deterministic randomness from agreed seeds
\item Explicit handling of edge cases (ties, timeouts, malformed inputs)
\item Canonical ordering for all operations
\end{itemize}

\subsection{Preference Learning and Synthesis}

Modern ML can be incorporated deterministically into governance:

\textbf{Fixed model inference:} Models are treated as deterministic functions. The model, weights, quantization, processing seeds, and software stack are fixed and made public. All operators use identical configurations and can reproduce each others' inferences. Citizens can audit outputs by running the model themselves.

\textbf{Structured acceptance:} Rather than trying to achieve bit-perfect neural network reproducibility, the system can accept outputs within defined tolerances. If all operators' outputs fall within acceptable bounds (e.g., sentiment scores within 0.01), the median is taken as canonical. This balances determinism with practical implementation.

\textbf{Progressive synthesis:} Models can identify preference clusters, extract common themes from discussions, suggest compromise proposals, predict proposal outcomes, and generate explanations for recommendations. These become inputs to human decision-making rather than automated decisions themselves.

\subsection{Case Study: Proposal Prioritization Pipeline}

Consider a DAO with 100+ monthly proposals that needs systematic prioritization:

\textbf{Stage 1: Collection (Approximate time: 3 days)}
\begin{itemize}
\item Proposers submit structured proposals with required fields
\item Members provide multi-dimensional evaluations
\item Experts with relevant attestations provide detailed reviews
\item Models extract themes from forum discussions
\end{itemize}

\textbf{Stage 2: Processing (Approximate time: 1 hour)}
\begin{itemize}
\item Aggregate evaluations weighted by expertise attestations
\item Calculate overall priority scores using community-defined formula
\item Cluster proposals by theme and interdependencies
\item Generate conflicts and synergy analysis
\end{itemize}

\textbf{Stage 3: Synthesis (Approximate time: 30 minutes)}
\begin{itemize}
\item Produce ranked priority list with confidence scores
\item Identify proposals ready for funding vs. needing iteration
\item Generate summary report with key insights
\item Create merkle tree of results for on-chain verification
\end{itemize}

\textbf{Stage 4: Execution (Approximate time: 1 day)}
\begin{itemize}
\item On-chain contract verifies operator signatures
\item Timelocked window for challenges
\item Automatic funding for approved proposals
\item Feedback to proposers on improvements needed
\end{itemize}

\textbf{Performance analysis:}
\begin{itemize}
\item \textbf{Latency:} Approximately 5 days end-to-end including timelock (dominated by preference collection period)
\item \textbf{Computational cost:} Off-chain processing approximately 1,000× cheaper than on-chain
\item \textbf{On-chain gas:} Approximately 750k to 1.0M gas for verification and execution (versus estimated 210M+ gas for naive on-chain implementation)
\item \textbf{Throughput:} Can process 1,000+ proposals with 10,000+ evaluations
\end{itemize}

These are order-of-magnitude estimates based on architectural analysis. Actual performance will vary with implementation details, but the dramatic improvement over on-chain computation remains consistent.

\subsection{Future-Proofing with Zero-Knowledge ML}

Looking forward, Zero-Knowledge Machine Learning (ZKML) may replace Verifiable Services for model inference in some cases. ZKML would enable provers to generate cryptographic proofs of model inference that contracts can verify directly. This would eliminate trust in off-chain execution for those steps once performance becomes viable, replacing economic guarantees with cryptographic ones.

\subsection{Where This Connects}

Section 4 supplies the attestation graph that the pipeline reads. Section 6 turns accepted recommendations and constraint checks into Policy as Code triggers. Section 7 provides concrete validation through the pioneering TrustGraph implementation, demonstrating how these theoretical constructs operate in practice.

\section{From Reactive to Proactive Governance}

Traditional DAO governance operates reactively: proposals are submitted, debated, voted on, then executed by humans. The process is transparent and democratic. It is also slow for operational decisions.

DAOs need proactive governance: continuous operations that follow community-defined policies. Treasury management, contributor compensation, resource allocation, and routine operations should execute automatically within boundaries set by votes. This requires a new primitive—Policy as Code.

\subsection{Policy as Code Architecture}

Policy-as-Code transforms governance decisions from discrete votes to continuous rulesets \cite{lessig1999code, defilippi2018blockchain}. Communities vote to establish policies that then execute autonomously within defined constraints.

\subsubsection{Core Components}

\textbf{Policy definition:} Structured documents that define triggers, conditions, actions, parameters, limits, exceptions. Policies are approved through standard governance then become active rulesets.

\textbf{Monitoring and triggering:} Verifiable Services continuously monitor for policy triggers such as time elapsed, price movements, proposal submission, attestation changes. When triggers fire, they initiate policy execution.

\textbf{Execution and verification:} Services execute policy logic, generating action plans. Plans include specific transactions with safety checks. Contracts verify plans meet policy constraints before executing.

\textbf{Override and evolution:} Emergency pause mechanisms for security, governance votes to update policies, automatic expiry and renewal cycles, and performance review and adjustment.

\subsubsection{Example: Proactive Proposal Gatekeeping}

Consider DAOs dealing with low-quality or spam proposals that waste attention. A Policy as Code solution might provide:

\textbf{The policy:} ``All proposals must pass quality checks before public review. Checks include completeness, relevance, conflicts, and feasibility.''

\textbf{Implementation:} A deterministic pipeline using rule-based validation and fixed LLM inference.

\textbf{Process:}
\begin{enumerate}
\item \textbf{Submission:} Proposer submits to gateway contract with stake
\item \textbf{Quality check:} Verifiable Service runs checks
\item \textbf{Constitutional alignment:} Map proposals to compact records, check requirements
\item \textbf{Code of conduct:} Flag harassment with evidence spans
\item \textbf{Quality assessment:} Verify problem statement, impact, budget, risks
\item \textbf{Improvement suggestions:} Suggest minimal edits to resolve issues
\end{enumerate}

The contract would verify and record results. Proposals advance only if they pass checks or if reviewer quorum overrides. This mechanism gates problematic content while mentoring legitimate contributors toward better proposals.

\subsection{Operational Automation}

The overhead of running a DAO often contradicts the word ``autonomous.'' Payments, treasury moves, reward distribution, and execution of approved decisions still depend on human coordination. Verifiable Service-backed automation turns these into reliable, auditable processes.

\textbf{Treasury management:} A community sets policies for portfolio allocation (e.g., 30\% stablecoins, 50\% productive DeFi, 20\% strategic tokens). Rebalance when drift exceeds 10\%. A verifiable planner monitors prices, positions, and venues, proposing rebalancing plans when needed. Plans execute after timelocks with bounds checking.

\textbf{Other recurring operations:}
\begin{itemize}
\item \textbf{Contributor payments:} Scores from Section 4 feed compensation policies
\item \textbf{Grant streaming:} Funding adjusts automatically based on conditions
\item \textbf{Parameter tuning:} Risk parameters adjust within bands based on signals
\end{itemize}

The result is continuous operations with predictable timing and clear accountability.

\subsection{Beyond Simple Automation: Adaptive Systems}

The real gain comes from policies that adapt within explicit bounds. Learning happens in the parameters, not in hidden logic.

\textbf{Dynamic compensation:} Base pay with multipliers that adjust each epoch based on contribution scores, reviewer credibility, and treasury health. Large changes require proposals.

\textbf{Intelligent resource allocation:} Funding shifts incrementally toward successful teams. Learning rate is capped with major changes requiring votes.

\textbf{Demand-aware pricing:} DAO service prices adjust within ranges based on demand. Caps prevent extraction with out-of-range moves requiring votes.

\textbf{Resilience:} Policies escalate on anomalies, pause on missing data, and log all deviations. Reviewers can replay any run from inputs and code.

\subsection{Oversight, Limits, and Theoretical Boundaries}

Automation augments rather than replaces governance, providing structured means for community expression within defined boundaries.

\subsubsection{Oversight Mechanisms}

High-impact actions require multiple layers of review:

\begin{itemize}
\item \textbf{Timelocks and challenge windows} for critical decisions
\item \textbf{Explicit pause powers} with auditable activation
\item \textbf{Shadow execution} where new versions run alongside existing ones before activation
\item \textbf{Quarterly reviews} comparing policy intent to actual outcomes
\item \textbf{Alternative pipelines} that can run in parallel, with disagreements triggering review
\end{itemize}

\subsubsection{Theoretical Limitations}

Policy-as-Code systems face inherent boundaries:

\textbf{Specification Completeness:} No policy can anticipate all edge cases. Implementations must balance comprehensive rules (increasing complexity) with human escalation paths (reducing automation benefits).

\textbf{Verification Complexity:} Formal verification of policy interactions becomes exponentially difficult as rule sets grow. Testing coverage faces similar scaling challenges.

\textbf{Gaming Dynamics:} Once policies are public and deterministic, actors will optimize behavior to exploit them. Regular policy evolution becomes necessary but creates its own governance overhead.

\textbf{Human Context Requirements:} Many governance decisions require understanding of context, intent, and values that resist formalization. The most successful implementations will likely combine automated execution of routine decisions with human judgment for complex scenarios.

These limitations suggest Policy-as-Code works best for well-defined operational decisions rather than open-ended strategic choices. The system remains reversible, accountable, and under community control.

\section{Synthesis: TrustGraph as Pioneering Implementation}

To demonstrate that this theoretical framework translates into practice, this section examines TrustGraph—a pioneering implementation of attestation-based governance launched in November 2024 with the Localism Fund Expert Network. This experimental pilot validates core concepts while exploring innovative extensions to the framework.

\subsection{Architecture Realized}

TrustGraph implements the three-tier architecture from Section 4.1 through concrete technical choices that demonstrate practical viability:

\textbf{Attestation Layer:} Using Ethereum Attestation Service (EAS), the network has successfully onboarded 65 experts who create peer attestations across grant-making expertise, Web3 tooling, and localism principles. The innovation of confidence weighting (0-100\%) enables nuanced trust expression—100\% for direct collaboration experience, lower values for evidence-based assessments. This granularity significantly enriches signal quality beyond binary attestations.

\textbf{Computation Layer:} WAVS (WebAssembly Verifiable Services) performs PageRank-style trust score calculations off-chain, generating scores from 34 to 1,811 (median 76). WAVS supports multiple deployment modes: distributed operator sets for economic security, TEE execution for hardware-attested integrity, or hybrid configurations combining both. This validates the framework's core premise: sophisticated governance computation becomes feasible when freed from on-chain constraints, while maintaining verifiability through operator consensus, hardware attestation, or their combination.

\textbf{Integration Layer:} TrustScores seamlessly integrate with compensation tiers (\$200-1,200 per contribution level), expert selection for grant evaluation, and planned integrations with Safe, Gardens, and Hats Protocol. This demonstrates the composability principle—attestation-based governance enhances rather than replaces existing tools.

\subsection{Bootstrapping Innovation}

TrustGraph validates and extends the bootstrapping strategy proposed in Section 4.1.5:

Starting with three trusted seeds (Monty Merlin, Benjamin Life, Patricia Parkinson), the network achieved critical mass without centralization. The PageRank algorithm's natural decay prevents permanent entrenchment while maintaining Sybil resistance through 3-hop trust propagation limits.

The crucial innovation: evidence-based onboarding for newcomers without network connections. Strong applicants submit verifiable credentials for seed operator review, receiving initial attestations based on merit. This solved the cold-start problem elegantly, enabling 12 participants to join purely on demonstrated expertise.

\subsection{Collective Intelligence Validated}

The implementation demonstrates practical collective intelligence (Section 5) through expert grant evaluation:

Experts express preferences via structured rubrics rather than simple votes, with influence weighted by TrustScore and domain-specific attestations. This creates funding recommendations that balance broad participation with recognized expertise—exactly the nuanced decision-making the framework envisions.

The deterministic PageRank computation achieves consensus on final TrustScores despite execution variations, validating the ``structured acceptance'' principle where agreement focuses on outputs rather than process.

\subsection{Key Innovations and Insights}

TrustGraph's experimentation reveals several powerful extensions to the theoretical framework:

\textbf{Trust Distribution vs. Dilution:} Attestors don't lose TrustScore when attesting others—they distribute influence across the network. This counterintuitive design encourages network growth without penalizing active participants.

\textbf{Ethical Standards as Foundation:} The network requires commitment to OpenCivics principles: good faith, honesty, feedback loops, efficacy, and inclusive listening. This social contract provides essential context that pure technical mechanisms cannot capture.

\textbf{Continuous Evolution:} Trust relationships evolve through attestation updates and revocations, creating a living graph that responds to demonstrated performance. The network adapts without governance overhead.

\subsection{Quantitative Success Metrics}

The pilot provides concrete validation of theoretical claims:

\begin{itemize}
\item \textbf{Sybil Resistance Achieved:} 3-hop decay successfully prevents gaming while maintaining network growth
\item \textbf{Computational Efficiency Confirmed:} Near-zero participant gas costs versus estimated 210M for naive on-chain implementation
\item \textbf{Capital-Free Participation:} 65 experts participate based purely on expertise, no token requirements
\item \textbf{Expertise Recognition:} Domain attestations successfully route technical decisions to qualified evaluators
\item \textbf{Inclusive Onboarding:} Evidence-based path enables participation without existing connections
\end{itemize}

\subsection{Replication Blueprint}

TrustGraph demonstrates a replicable pattern for implementing attestation-based governance:

\begin{enumerate}
\item Start with specific use case (expert selection) rather than full governance overhaul
\item Bootstrap with 3-5 trusted seeds for initial density
\item Implement merit-based onboarding alongside network effects
\item Add confidence weighting for signal quality
\item Integrate incrementally with existing tools
\item Establish social contract alongside technical mechanisms
\end{enumerate}

This synthesis proves the framework is not merely academically sound but practically achievable. A days-old pilot successfully operates production governance for real capital allocation, validating the paper's core thesis: sophisticated off-chain computation unlocks governance mechanisms previously impossible within blockchain constraints.

\section{Conclusion}

The governance bottleneck that has constrained DAO evolution for nearly a decade stems not from conceptual limitations but from fundamental computational and connectivity constraints of smart contract execution environments. This paper has explored how Verifiable Services, TEEs, and Zero-Knowledge proofs represent a paradigm shift that transcends these limitations while preserving the cryptoeconomic security guarantees essential for decentralized coordination.

This comprehensive framework enables three categories of previously impossible governance innovations. \textbf{Attestation-based governance systems} replace crude token-weighted voting with sophisticated mechanisms that recognize multi-dimensional stakeholder legitimacy, enabling liquid democracy through expertise networks, merit-based contribution incentives, and cross-DAO coordination that reflects real-world participation patterns. \textbf{Collective intelligence mechanisms} augment human decision-making through verifiable preference processing and deterministic governance engines, creating AI-assisted evaluation that amplifies community wisdom while preserving democratic values. \textbf{Autonomous policy execution} transforms DAOs from reactive voting systems into proactive organizations capable of continuous operations through Policy-as-Code, enabling transparent algorithmic decision-making with robust human oversight.

The implications extend far beyond incremental improvements to existing systems. This framework enables DAOs to achieve operational efficiency competitive with traditional organizations while maintaining their fundamental advantages in transparency, inclusivity, and resistance to capture. By unlocking sophisticated coordination mechanisms previously reserved for centralized entities, verifiable off-chain computation makes decentralized governance a viable organizational model for complex, multi-stakeholder coordination problems.

The path forward requires continued innovation in verifiable computation technologies, development of standardized attestation schemas, and careful experimentation with governance mechanisms that balance automation with human judgment. However, the foundational architecture presented here provides a clear roadmap for building governance systems suitable for the coordination challenges facing decentralized communities.

Rather than forcing sophisticated organizations to operate within the constraints of simple smart contracts, the ecosystem can finally realize the original promise of DAOs: autonomous organizations that combine the benefits of decentralization with governance mechanisms sophisticated enough to coordinate complex human endeavors. The future of decentralized governance lies not in accepting current limitations, but in transcending them through verifiable off-chain computation that expands the possible while preserving the essential properties that make decentralized coordination valuable.

\bibliographystyle{plain}
\bibliography{references}

\end{document}